\documentclass[12pt]{article}
\usepackage{graphicx}

\title{Generative Adversarial Networks for Model Order Reduction in Seismic Full-Waveform Inversion}
\author{Alan Richardson (Ausar Geophysical)}
\begin{document}
\maketitle
\begin{abstract}
I train a Generative Adversarial Network to produce realistic seismic wave speed models. I integrate the generator network into seismic Full-Waveform Inversion to reduce the number of model parameters and restrict the inverted models to only those that are plausible. Applying the method to a 2D section of the SEAM Phase I model, I demonstrate that it can produce more plausible results than conventional Full-Waveform Inversion.
\end{abstract}
\section{Introduction}

The concept of a Generative Adversarial Network (GAN) was recently proposed as a method for producing new samples that appear similar to those in a given dataset \cite{goodfellow2014generative}. It consists of two neural networks: a generator that attempts to transform an input latent vector into a realistic sample, and a discriminator that attempts to identify fake examples. When trained using seismic velocity models, the generator learns to produce plausible velocity models from latent vectors. As the dimension of the latent vector space is smaller than that of the model space, this can be considered to be a form of model order reduction. As the generator is differentiable, gradients may be backpropagated through it, allowing the latent vector to be optimized.

My hypothesis is that a GAN trained to produce realistic wave speed models, from latent vectors with fewer parameters than the full models, can be used to reduce the number of parameters to be inverted in seismic Full-Waveform Inversion (FWI). Fewer parameters should make the method less prone to overfitting, as this reduces the flexibility of the model, constraining it to be realistic. Using a GAN in this way could therefore be thought of as a form of regularization. A regularizer that favors plausible features, such as deformed layers of constant velocity and salt bodies, is difficult to express mathematically, but the GAN constructs one automatically. Reducing the number of model parameters should also make using stochastic optimization to find an initial model more computationally feasible.

\section{Materials and Methods}

The proposed method consists of two components: a GAN that generates realistic seismic velocity models, and a modified FWI implementation that incorporates the GAN.

\subsection{Generative Adversarial Network}
To evaluate the method, I test it on a small 2D model, so a GAN generator that produces 2D samples is needed. The DCGAN \cite{radford2015unsupervised} network structure has been shown to be successful for producing realistic 2D images. This consists of a latent vector with 100 elements, and five convolutional layers in both the generator and discriminator (convolution transpose for the generator). I use it with only one modification: I add a term to the cost function of the generator that penalizes wave speeds outside the range 1450 -- 5000 m/s. In my implementation, this cost is equal to the magnitude of the deviation from this range.

Training the GAN requires many velocity model samples. To produce these, I use a simple code to generate models with $64 \times 64$ cells that consist of sedimentary layers and salt bodies. The sedimentary layers are constant-velocity layers, with the velocity generally increasing with depth, that are randomly distorted. To produce the salt body that is added to each model, I randomly distort a circle, and fill it with a constant velocity of approximately 4600 m/s. I use $2^{17} = 131072$ of these models, and train for eight epochs with the Adam optimizer using a batch size of 64 and a learning rate of 0.0001.

\subsection{Full-Waveform Inversion}
In conventional FWI, the residual between the true data and data produced by forward propagating waves through a model, is backpropagated through the model to calculate the gradient of the cost function with respect to the model parameters. This gradient is then used to update the model parameters. To incorporate the GAN, I modify this last step. Instead of updating the wave speed model directly, I further backpropagate the gradient through the GAN to update the latent vector that produced the model. The models considered by FWI are therefore limited to those that can be produced by the generator, ensuring that they are always plausible, and reducing the number of parameters to the dimension of the generator's latent vector space.

FWI requires an initial model from where it begins iteratively converging toward the true model. In the proposed method, the latent vector that produces this initial model must first be found. I use a random search to find the initial latent vector. To do this, I create random latent vectors and forward propagate through the resulting models to calculate the value of the FWI cost function associated with each. I scale the data residual by time to approximately compensate for spreading losses. To reduce computational cost, I sum all of the shots into a single shot with multiple sources. I use the vector with the lowest cost function value as the initial latent vector for gradient-based optimization.

\subsection{Method overview}
For clarity, the following describes the steps of the proposed method:
\begin{enumerate}
        \item Train the GAN using example seismic models
        \item Find the initial latent vector
        \begin{enumerate}
                \item Initialize the latent vector with random numbers
                \item Apply the generator to the vector to create a model
                \item Forward propagate sources through the model to calculate its FWI cost function value
                \item Iterate until stopping criterion, and use latent vector with lowest cost
        \end{enumerate}
        \item Iteratively optimize the latent vector
        \begin{enumerate}
                \item Apply the generator to the vector to create a model
                \item Perform one iteration of FWI to calculate the gradient of the cost function with respect to the model
                \item Backpropagate the gradient through the generator to update the latent vector
                \item Iterate until stopping criterion
        \end{enumerate}
\end{enumerate}

\subsection{SEAM dataset}
To test the ability of the method to perform seismic inversion, I use a model derived from the SEAM Phase I model~\cite{fehler2011seam}. I use a 2D section of the Vp model extracted from the 23900 m North line, covering from 9600 m to 16000 m in the horizontal direction and from 100 m to 6500 m in the depth direction. The extracted model is interpolated onto a grid with 100 m cell spacing (making it $64 \times 64$ cells). The dataset contains shots forward modeled on this model, using a 1 Hz Ricker wavelet as the source with a source spacing of 500 m and receiver spacing of 100 m along the top surface.

For conventional FWI, I use 20 LBFGS steps, each requiring that the cost and gradient of the entire dataset are calculated 20 times. With 13 shots, and the equivalent of two forward modeling steps to calculate the cost and gradient, this equates to $20 \times 20 \times 13 \times 2 = 10400$ forward shot modeling steps. I use a learning rate of 0.1 as, unlike the proposed method, large step sizes can lead to unrealistically high wave speeds that require the wave propagator to use a small time step size. I start from an initial model that increases by 0.5 m/s/m with depth from 1490 m/s at the surface.

For the proposed method, I use 50 vectors in the random search for an initial latent vector. I then run 5 LBFGS steps with a learning rate of 1 to optimize this result. The combination of these involves $50 + 5 * 20 * 13 * 2 = 2650$ forward shot modeling steps.

\section{Results}

I present results from the two stages of the method: training the GAN to produce realistic seismic models, and using the resulting generator during FWI to invert for a model.

The code to reproduce these results is included in the ancillary files accompanying this article.

\subsection{GAN training}
\begin{figure}
        \includegraphics{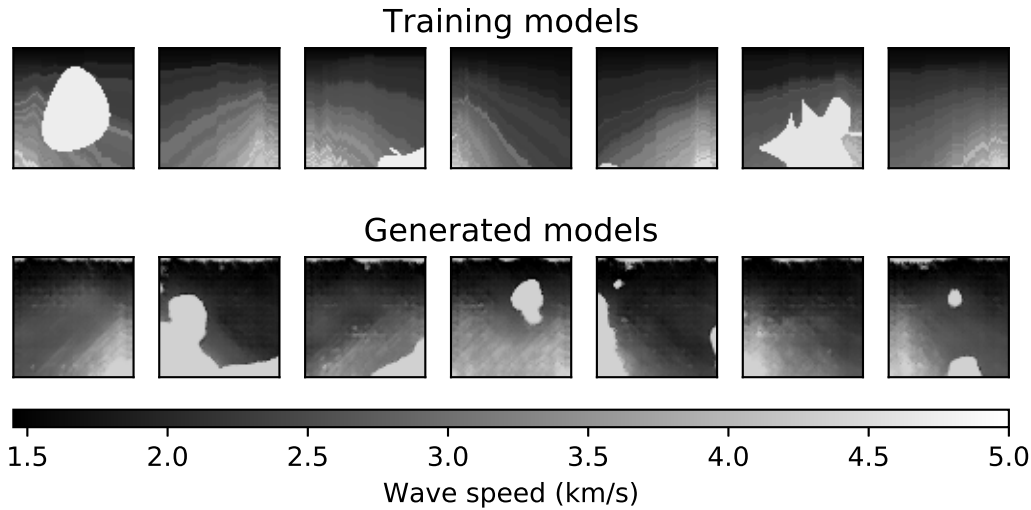}
        \caption{Example models from the GAN training dataset and produced by the generator after training. The generated models look similar to the training models.}
        \label{fig:gan_models}
\end{figure}
A random selection of models used to train the GAN, and output models produced by the generator network from random latent vectors, are shown in Figure \ref{fig:gan_models}. The generator appears to have learned to produce realistic seismic models, as the generated models are similar to the training models.

\subsection{FWI}
\begin{figure}
        \includegraphics{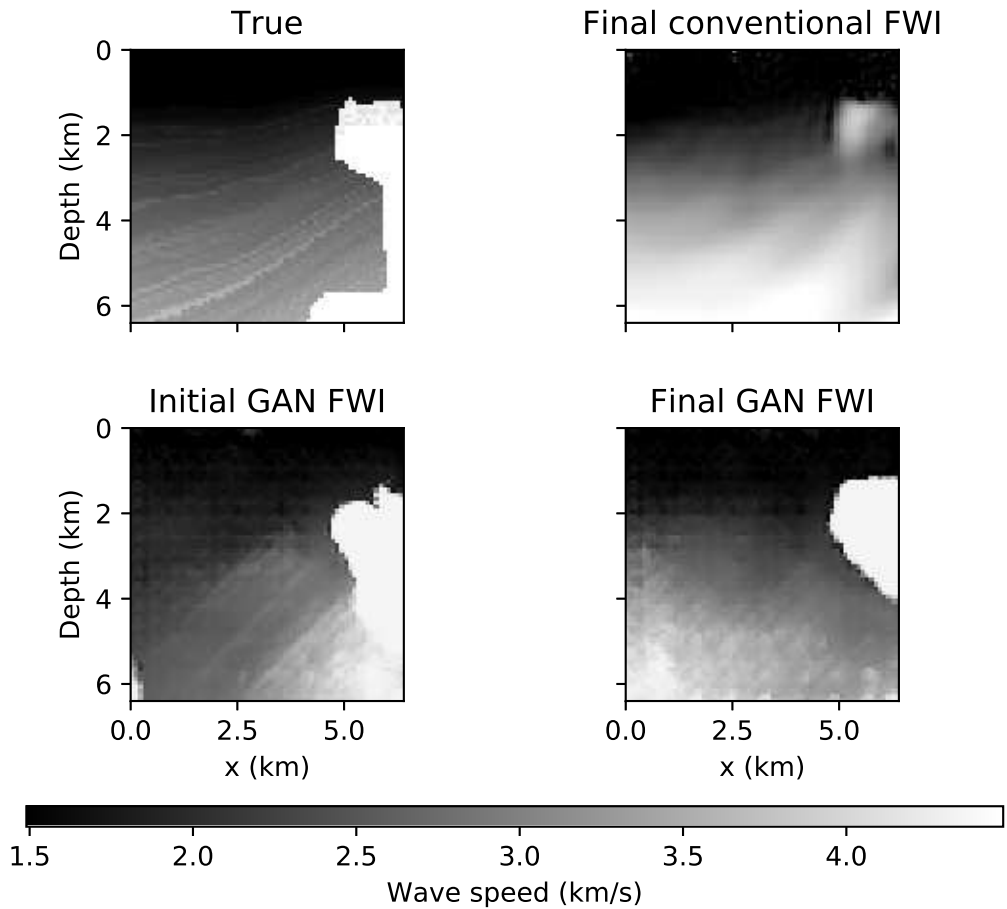}
        \caption{The true model, conventional FWI result, initial GAN model, and optimized GAN model.}
        \label{fig:fwi}
\end{figure}
The results of conventional FWI and the proposed method are presented in Figure \ref{fig:fwi}. The conventional FWI result is quite smooth, as expected. It also contains only the top portion of the salt body, which then fades to the background model with depth. More of the salt may have been captured by using modifications such as those proposed by \cite{esser2018total}. The initial model for the proposed method, found after a random search, is already quite close to the true model. Iterative optimization improves the result further. The result looks plausible. However, compared to the true model, the shape of the salt body is not fully correct, and the lowest portion of it is missing. The structure of the sedimentary layers also appears to be less accurate than in the conventional result, but the wave speed is closer to the truth. One may argue that, with its sharp salt body edges, the proposed method favors precision over accuracy, but the result is closer to the true model than that found by conventional FWI, with RMS model errors of 510 m/s and 758 m/s respectively.

\section{Discussion}

\paragraph{GAN training and plausible models} The generator network learns to produce models similar to the input training models during GAN training. The plausibility of the generated models thus depends on that of the training models. I use simple synthetic models during training. It is likely that better results would have been obtained from the modified FWI proposal if more realistic models were used for training. In addition to only producing plausible models, another concern is whether the generator can produce all possible plausible models. GAN training is notoriously susceptible to problems such as collapsing to a state where the generator always produces the same output. Care (and luck) is therefore needed to ensure that GAN training is successful.

\paragraph{Computational cost} The proposed method has three differences from conventional FWI: the addition of a generator, a different method of finding an initial model, and the need to train the GAN. Forward modeling through the generator network to produce a model and backpropagating through it to update the latent vector are of negligible computational cost compared to wave propagation through the model. Adding the generator therefore does not noticeably increase the cost of FWI. In the results above, I find an initial latent vector using a random search. The computational cost of this depends on the number of random vectors that are evaluated, which I expect would need to be higher for more realistically sized models. Other initialization methods are available, as I discuss below. Using a GPU, training the GAN takes about the same amount of time as the FWI step for the results above. Once the generator is trained, however, it can be used for any dataset with the same model size. This is discussed further below.

\paragraph{Initial model} If a good initial model is available, such as one derived from tomography, then it may be used instead of finding one with a random search. A latent vector that approximately produces the chosen initial model must be found. This can be achieved by starting with a random latent vector and using the mean squared error between the model generated from this and the desired initial model to iteratively optimize the latent vector. An outline of this approach follows.
\begin{enumerate}
        \item Initialize the latent vector with random numbers
        \item Iteratively optimize the vector
        \begin{enumerate}
                \item Apply the generator to the vector to create a model
                \item Calculate the residual between the generated model and the desired initial model
                \item Backpropagate through the generator to update the latent vector
                \item Iterate until stopping criterion
        \end{enumerate}
\end{enumerate}

\paragraph{Generating different model sizes} In my implementation, the generator network produces a $64 \times 64$ array. This suggests that a new generator would need to be trained if another model size is desired. It is quite likely, however, that the models it produces would still look plausible even if they were stretched, for example to produce a $64 \times 128$ model. Stretching the outputs in this way would allow the same generator to be used for a variety of model sizes, as long as the model gradient is compressed back to $64 \times 64$ so that it can be backpropagated through the generator to update the latent vector during FWI. It may also be possible to stitch together multiple generated models to produce a larger model.

\paragraph{Local minima} Conventional FWI suffers from a problem with local minima, and so a good initial model is often required to ensure convergence to a satisfactory result. I expect the proposed method to be similarly dependent on a good initial model. It is even possible that this dependence may be stronger. I tried starting the proposed FWI method from a random initial latent vector, but it did not appear to converge toward the true model, for example. This problem may be mitigated by the reduced cost of randomly searching for an initial model.

\paragraph{Related work} The most closely related work appears to be \cite{mosser2018rapid}, in which the authors use GANs to transform a seismic image into a seismic wave speed model. As a wave speed model is needed to create a seismic image, this method may be applicable later in the processing workflow than my proposal. In another use of GANs for seismic applications, the authors of \cite{siahkoohiseismic} use one to replace missing seismic data. Although GANs are not used, \cite{lewis2017deep} has similarities as it uses deep neural networks during FWI, identifying areas likely to contain salt.

\section{Conclusion}
The hypothesis appears to be true. It is possible to train a GAN that generates plausible seismic models. This generator can be used in a random search to quickly find a good initial latent vector. Optimizing this vector, by combining conventional FWI with the generator, produces a realistic result that may be a good starting model for further refinement with conventional FWI.

\bibliography{seismic_modelgen}{}
\bibliographystyle{plain}
\end{document}